\documentclass[aps,prl,twocolumn]{revtex4-2}
\usepackage{times,amsmath,amssymb,amsthm,amsfonts,bm,braket,dcolumn,graphicx,mathrsfs,multirow,physics,xcolor,txfonts}
\usepackage{soul}
\usepackage[colorlinks=true, linkcolor=blue!75!black, citecolor=blue!75!black, urlcolor=blue!75!black]{hyperref}
\renewcommand{\Re}{\operatorname{Re}}

\newcommand{\llangle}{\langle\!\langle}
\newcommand{\rrangle}{\rangle\!\rangle}
\newcommand{\clr}{\color{red!75!black}}

\newtheorem{theorem}{Theorem}

\begin{document}

\title{Dynamical Quantum Phase Transitions and Many-Body Backflow in Open Quantum Systems}

\author{Kai Zhang}
\thanks{These authors contributed equally}
\affiliation{Department of Physics, University of Michigan Ann Arbor, Ann Arbor, Michigan, 48109, United States}
\author{Chang Shu}
\thanks{These authors contributed equally}
\affiliation{Department of Physics, University of Michigan Ann Arbor, Ann Arbor, Michigan, 48109, United States}
\author{Kai Sun}
\email{sunkai@umich.edu}
\affiliation{Department of Physics, University of Michigan Ann Arbor, Ann Arbor, Michigan, 48109, United States}

\begin{abstract}
Dynamical quantum phase transitions (DQPTs) are non-equilibrium transitions characterized by the orthogonality between an initial quantum state and its time-evolved counterpart following a sudden quench. Recently, studies of this phenomenon have been extended beyond closed quantum systems to include environmental interactions, often modeled through non-Hermitian effects. However, because non-Hermitian descriptions neglect both quantum jump processes and interaction effects, the ultimate fate of interacting quantum systems under full open-system quantum dynamics remains an open question. In this Letter, by incorporating both interactions and full Liouvillian dynamics, we prove that DQPTs in open quantum systems remain robust when subject to either particle loss or particle gain alone, but are generically smeared out when both processes coexist, as a result of many-body particle backflow. Furthermore, we uncover a non-perturbative dynamical effect: even a weak admixture of gain (loss) into a system with loss (gain) can dramatically reshape the long-time behavior of DQPT dynamics, leading to substantial deviations over time. These phenomena—including the universal smearing of DQPTs and the emergence of large dynamical deviations in the long-time limit—arise intrinsically from non-equilibrium many-body effects in open quantum systems. Our findings are general and substantiated by both analytical arguments and numerical simulations.
\end{abstract}

\maketitle

\emph{{\clr Introduction}.---}~Analogous to the nonanalytic behavior observed in equilibrium quantum and classical phase transitions~\cite{Sachdev1999}, a counterpart has been identified in the real-time evolution of quantum systems, known as the dynamical quantum phase transition (DQPT)~\cite{Heyl2018Review}. 
In non-equilibrium many-body systems, DQPT arises when the system is quenched across an equilibrium phase transition, for instance, in the transverse-field Ising model, where the system is suddenly quenched from a ferromagnetic to a paramagnetic phase~\cite{Heyl2013PRL}. 
Conceptually, DQPT occurs at a critical time $t_c$ when the time-evolved wavefunction becomes orthogonal to the initial state. 
This is typically expressed as $\langle \Psi_0| U(t_c) | \Psi_0\rangle = 0$, where $U(t)$ is the time-evolution operator corresponding to the quenched Hamiltonian. 
Following the theoretical~\cite{Fischer2006PRL,Fischer2007PRL,Heyl2015PRL,Vajna2015PRB,Zvyagin2016Review,Dutta2017PRB,Heyl2018Review,ShuChen2025PRB} and experimental progress~\cite{DQPTObservation2017,Monroe2017Nature,Weitenberg2018NatPhys,XuePeng2019PRL,FanHeng2019PRA,DuanLM2020PRL}, it was further revealed that DQPTs can also occur when the underlying equilibrium system undergoes a topological phase transition, thereby establishing a deeper connection between dynamical and topological phases~\cite{Vajna2015PRB,Budich2016PRB,Huang2016PRL,Budich2017PRB,Dutta2017PRB,YiWei2018PRA,ShuChen2025arXiv}.

However, in realistic quantum devices, the system's dynamics is inevitably affected by the external environment, such as dephasing or dissipative effect. It is therefore essential to examine the DQPTs in open quantum systems. 
Recent advances in quantum platforms, such as ultracold atoms~\cite{GBJo2025Nature}, trapped ion systems~\cite{ZhangWei2021PRL}, and photonic quantum walks~\cite{XuePeng2020,Xue2025LSA}, have enabled unprecedented precise control over non-unitary quantum dynamics.
Unlike closed quantum systems, in which the system evolves under unitary time evolution governed by the {Schr\"odinger} equation, open quantum systems require a more general framework that tracks the evolution of the density matrix $\rho$. 
With the Born-Markov approximation of environment, the system's dynamics is governed by the Lindblad master equation~\cite{Breuer2002book,Daniel2020Review},
\begin{equation}\label{eq_Lindblad}
    \frac{d\rho}{dt} = -i[H, \rho] + \sum\nolimits_{\mu} \left(L_{\mu} \rho L_{\mu}^\dagger - \frac{1}{2}\{L^{\dagger}_{\mu}L_{\mu}, \rho\} \right),
\end{equation}
where $H$ represents the system's Hamiltonian, and $L_{\mu} \rho L_{\mu}^\dagger$ corresponds to quantum jump terms. 
When the quantum jump processes can be neglected over short timescales, the system's dynamics can be approximated by the non-Hermitian Hamiltonian~\cite{QuantumTrajectory1992PRL,Ashida2020,Ding2022NRP,Bergholtz2021RMP}: $d\rho/dt = -i (H_{\text{eff}}\rho - \rho H^{\dagger}_{\text{eff}})$ with $H_{\text{eff}} = H - \frac{i}{2}\sum_{\mu} L_{\mu}^\dagger L_{\mu}$. 
Motivated by a wide range of striking phenomena enabled by non-Hermiticity, such as exceptional points~\cite{Heiss2012,YiWei2024Review}, non-Hermitian skin effect~\cite{Yao2018,SongFei2019}, and the expansion of topological phases beyond Hermitian constraints~\cite{SatoPRX2019}, non-Hermitian physics has received growing attention in recent years. 

More recently, DQPTs in non-Hermitian systems have attracted growing interest, since non-Hermitian Hamiltonians can exhibit a richer variety of topological phases~\cite{ZhouLW2018PRA,YiWei2019Review,XuePeng2019PRL,YiWei2021PRR,Sim2023PRL,JingYC2024PRL,ZhangL2025} and exotic phase transitions~\cite{Moessner2019NC,XP2021PRXQuantum,LiFX2024,Keshav2024PRA}. 
However, several fundamental questions remain to be clarified:
(i)~Whether DQPTs predicted by an effective non-Hermitian Hamiltonian can survive under a genuine open quantum dynamics? 
(ii)~If not, what is the correct criterion for identifying DQPTs in open systems governed by a Lindblad master equation? 
(iii)~Are there dynamical phenomena that go beyond the effective non-Hermitian description? 
Although some previous studies have extended DQPTs to open quantum systems, most of them focus on purely dissipative dynamics or specific examples~\cite{Valentin2020PRL,Dutta2018SciRep,FanHeng2018PRB,XuePeng2019PRL}, making general answers to the above questions remain unclear. 

In this Letter, we use the quantity $\tr[\rho_0 \rho(t)]$ as a probe for DQPTs in open quantum systems~\cite{Valentin2020PRL}, where $\rho_0$ and $\rho(t)$ denote the initial and final density matrix states evolved under Eq.~\eqref{eq_Lindblad}. 
The central result of this Letter is a theorem that establishes: 
(i) DQPTs can occur in open quantum systems with either particle loss or gain alone, where DQPTs are well captured by the corresponding non-Hermitian effective Hamiltonian;
(ii) Importantly, when both particle loss and gain are present, DQPTs are necessarily smeared out. This smearing arises from a many-body backflow effect in the double Hilbert space, which prevents the system from reaching the orthogonality condition essential for DQPT. 
Our theorem holds rigorously for momentum-conserving open quantum systems with interactions and arbitrary many-body gain and loss. 
It further applies to non-integrable many-body models, which we confirm through numerical simulations using the Monte-Carlo wavefunction method~\cite{Molmer1993}. 

Furthermore, we demonstrate that even a weak perturbation--introducing particle gain (loss) into a system with particle loss (gain)--can significantly impact the long-time dynamics of DQPT. 
This non-perturbative dynamical effect arises from the competition between the Liouvillian gap and the overlap of initial and steady states. 
Note that the DQPT indicator $\tr[\rho_0 \rho_t]$ is the quantum fidelity when the initial state $\rho_0$ is a pure state~\footnote{Quantum fidelity, defined as $F(\sigma, \rho) = \left( \tr \sqrt{ \sqrt{\sigma} \rho \sqrt{\sigma} } \right)^2$, quantifies the similarity between two density matrices $\sigma$ and $\rho$, and is experimentally accessible in quantum simulators. When either $\sigma$ or $\rho$ is a pure state, the fidelity simplifies to $\tr[\sigma \rho]$. In general, these two quantities satisfy the inequality $\tr[\sigma \rho] \leq F(\sigma, \rho)$.}. Hence, the theorem and the associated dynamical phenomena are directly accessible in quantum simulators such as ultracold atoms and trapped-ion platforms. 

\emph{{\clr Theorem: Many-body backflow smears out DQPTs}.---}~To prove the theorem, we first introduce the necessary preliminaries. 
Let the Hilbert space have dimension $D$, thus the Lindblad master equation evolves a $D \times D$ density matrix according to Eq.~\eqref{eq_Lindblad}, expressed as $d\rho/dt = \mathcal{L}(\rho)$, where $\mathcal{L}$ is the Liouvillian superoperator. 
To represent $\mathcal{L}$ as a matrix, we vectorize the density matrix by flattening it in a row order, mapping $|v_1\rangle\langle v_2| \rightarrow |v_1\rangle \otimes |v_2\rangle$. 
We denote the vectorized form of matrix $B$ as $|B\rrangle$, and define the inner product in this linear space by $\llangle A | B \rrangle = \tr(A^\dagger B)$. 
Using the identity $\text{vec}(ABC) \equiv (A \otimes C^T) |B\rrangle$, we obtain the matrix representation of the Liouvillian superoperator, $\mathbb{L} = - i (H_{\text{eff}} \otimes I - I \otimes H^{\ast}_{\text{eff}}) + \sum_{\mu} L_{\mu}\otimes L_{\mu}^{\ast}$ with $H_{\text{eff}} = H - \frac{i}{2}\sum_{\mu}L_\mu^\dagger L_\mu$. 
This procedure expresses $\mathcal{L}$ as a linear operator acting on the double Hilbert space $\mathcal{H} \otimes \mathcal{H}$. 
Throughout this Letter, we use double-ket notation to denote vectors in the double Hilbert space. 

Consider a quench process in a fermionic many-body lattice system with $N$ unit cells and $M$ orbitals per unit cell. 
Assume that the pre-quench Hamiltonian is Hermitian and that the system is initially prepared in its ground state with a fixed particle number $n_0$. The initial density matrix, $\rho_0 = |\text{GS}\rangle\langle \text{GS}|$, thus represents a pure state. 
At time $t = 0$, the system begins to evolve under the Lindblad master equation given in Eq.~\eqref{eq_Lindblad}. 
We consider the initial state that is a direct product over all momentum sectors labeled by $k$.
One can thus define the Loschmidt rate function as
\begin{equation}\label{eq_LoschmidtRate}
    \begin{split}
    G(t) &= \lim_{N\rightarrow +\infty} -\frac{1}{N} \ln \Pi_k g(k,t) \\ 
    &= \lim_{N\rightarrow +\infty} -\frac{1}{N} \ln \Pi_k \tr[\rho_0(k)\rho(k,t)].
    \end{split}
\end{equation}
DQPT occurs when $G(t)$ becomes nonanalytic at critical times $t_c$ in the thermodynamic limit $N \rightarrow +\infty$ (see Sup.~Mat.~I~\cite{SupMat} for details). 
Equivalently, DQPT occurs only if there exists a momentum $k_c$ and time $t_c$ such that $g(k_c,t_c) = 0$. 

Now we present the theorem. 
\begin{theorem}\label{thm_backflow}
    If the quenched Liouvillian superoperator includes only the loss (or gain) jump process, DQPTs can occur, and their emergence is determined by the effective non-Hermitian Hamiltonian. In contrast, introducing both gain and loss simultaneously{---}no matter how weak{---}inevitably smears out these dynamical phase transitions.
\end{theorem}
The loss ($\hat L_{\mu}^l$) and gain ($\hat L_{\mu}^g$) jump operators in the theorem are defined as follows. 
Suppose the effective Hamiltonian respects a $U(1)$ symmetry, i.e., $[\hat H_{\text{eff}}, \hat Q] = 0$.
This symmetry ensures that $\hat H_{\text{eff}}$ is block-diagonal, with each block labeled by a fixed charge sector $q$. 
The loss and gain dissipators are defined through the canonical commutation relation $[\hat Q, \hat L_{\mu}^{l/g}] = \mp m \hat L_{\mu}^{l/g}$, where $m>0$ is the amount by which the operator shifts the charge, and the $-$ ($+$) sign defines loss (gain) jump operator, respectively. 
In what follows, we consider the $U(1)$ symmetry as particle number conservation, and provide a proof of Theorem~\ref{thm_backflow}. 

\begin{figure}[t]
    \begin{centering}
    \includegraphics[width=.75\linewidth]{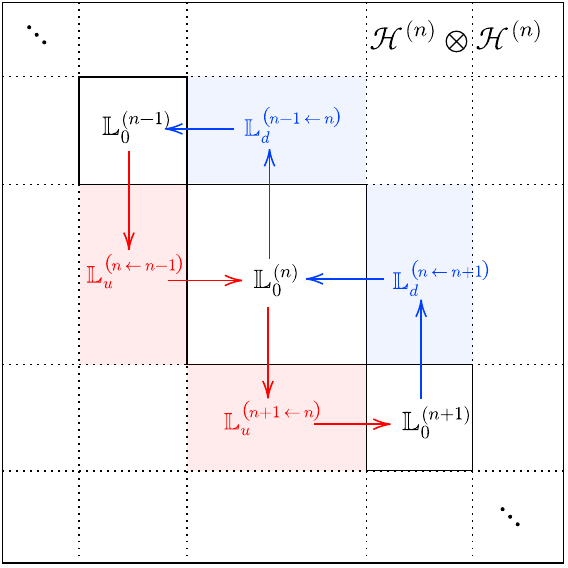}
    \par\end{centering}
    \protect\caption{\label{fig:1} Illustration of many-body backflow in the double Hilbert space. The system is initially prepared in the $n$-particle sector. Blue (red) arrows indicate transitions due to partial gain (loss) processes, which respectively increase (decrease) the particle number. When both are present, they create closed loops between different particle-number sectors, leading to many-body backflow. }
\end{figure}
 
\begin{proof}
    Consider a generic Liouvillian superoperator $\mathcal{L}$. 
    The particle loss and gain jump dissipators can be written as $\hat L^l = \sum_{s} \sqrt{\gamma_s^l}\hat c_1\dots \hat c_s$ and $\hat L^g = \sum_{s} \sqrt{\gamma_s^g}\hat c_1^{\dagger}\dots \hat c_s^{\dagger}$, respectively.
    The matrix representation of the Liouvillian in the double Hilbert space can then be decomposed into three components: $\mathbb{L} = \mathbb{L}_0 + \mathbb{L}_d + \mathbb{L}_u$. 
    The term $\mathbb{L}_0 = \bigoplus_{n=0}^M \mathbb{L}_0^{(n)}$ is block diagonal, as illustrated in Fig.~\ref{fig:1}, where each block $\mathbb{L}_0^{(n)}$ acts on the double Hilbert space $\mathcal{H}^{(n)}\otimes \mathcal{H}^{(n)}$ with fixed particle number $n$. 
    The diagonal term $\mathbb{L}_0$ describes the non-unitary evolution given by the effective non-Hermitian Hamiltonian $H_{\text{eff}}$ and takes the form $\mathbb{L}_0 = -i (H_{\text{eff}}\otimes I - I \otimes H^{\ast}_{\text{eff}})$.  
    The off-diagonal terms $\mathbb{L}_d$ and $\mathbb{L}_u$ correspond to upper and lower triangular structures, consist of blocks $\mathbb{L}_d^{(m \leftarrow n)}$ and $\mathbb{L}_u^{(n\leftarrow m)}$ with $n>m$, respectively, as shown in Fig.~\ref{fig:1}. 
    Note that off-diagonal blocks are contributed by the jump operators, and the superscript $({m\leftarrow n})$ corresponds to the shift from $n$- to $m$-particle sector. 
    
    We first consider the case involving only particle loss. 
    In the double Hilbert space, Liouvillian time evolution can be expanded into Dyson series as (see Appendix~A):
    \begin{equation*}
        e^{(\mathbb{L}_0 + \mathbb{L}_d)t} = e^{\mathbb{L}_0 t} \left( 1 + \int_0^t  d\tau e^{-\mathbb{L}_0 \tau} \mathbb{L}_d e^{\mathbb{L}_0 \tau} + \mathcal{O}(\mathbb{L}_d^2) \right),
    \end{equation*}
    where $\mathcal{O}(\mathbb{L}_d^2)$ represents higher-order terms. 
    DQPT is characterized by the quantity $g(t) = \tr[\rho_0\rho]$, or expressed as $\llangle \rho_0 |e^{(\mathbb{L}_0 + \mathbb{L}_d)t}|\rho_0 \rrangle$ in the double Hilbert space. 
    Since the initial state $|\rho_0\rrangle$ is assumed to have fixed particle number $n_0$, the evolution operator $e^{\mathbb{L}_0 t}$ preserves this sector, while operator $\mathbb{L}_d$ shifts it to sectors with fewer particles ($n_0\rightarrow m<n_0$), as indicated by the blue arrows in Fig.~\ref{fig:1}. 
    Consequently, all terms involving $\mathbb{L}_d$ yield zero overlap with $\llangle \rho_0 |$, thus, $g(t)$ with only particle loss reduces to:
    \begin{equation}\label{eq_lossjumpDQPT}
        \llangle \rho_0 |e^{(\mathbb{L}_0 + \mathbb{L}_d)t}|\rho_0 \rrangle = \llangle \rho_0 |e^{\mathbb{L}_0^{(n_0)} t} |\rho_0 \rrangle = \tr[\rho_0 e^{-iH_{\text{eff}}t}\rho_0 e^{iH^{\dagger}_{\text{eff}}t}],
    \end{equation}
    which is entirely governed by the effective non-Hermitian Hamiltonian.
    
    We now introduce particle gain as a perturbation with small strength ${0 < \lambda \ll 1}$~\footnote{${\lambda>0}$ is required by the fact that the probability of a jump process occurring must be positive.}~and prove that even an infinitesimal gain is sufficient to wash out the DQPTs. 
    Without loss of generality, we consider an $m$-particle gain dissipator: $L^{g} = \sqrt{\lambda}\hat c_1^\dagger \dots \hat c_m^\dagger$ and treat it as a perturbation. 
    Applying the Dyson series, the quantity $g(t)$ with both gain and loss can be expanded as (see detailed derivations in Appendix~A):
    \begin{widetext}
    \begin{equation}\label{eq_backflow}
    \begin{split}
        g(t) = \llangle \rho_0 | e^{(\mathbb{L}_0 + \mathbb{L}_d + \lambda \mathbb{L}_u)t} | \rho_0 \rrangle 
        = \llangle \rho_0 | e^{\mathbb{L}_0^{(n)} t} | \rho_0 \rrangle 
        + \lambda \left( 
        \int_0^t d\tau \int_0^{\tau} d\tau_1 \, 
        \llangle \rho^{(n)}_0 | 
        e^{\mathbb{L}^{(n)}_0(t-\tau)} \mathbb{L}^{(n\leftarrow n-m)}_u 
        e^{\mathbb{L}^{(n-m)}_0 (\tau-\tau_1)} \mathbb{L}^{(n-m\leftarrow n)}_d 
        e^{\mathbb{L}^{(n)}_0 \tau_1} | \rho^{(n)}_0 \rrangle \right. \\
        \left.
        + \int_0^t d\tau \int_0^{t-\tau} d\tau_1 \, 
        \llangle \rho^{(n)}_0 | 
        e^{\mathbb{L}^{(n)}_0(t-\tau-\tau_1)} \mathbb{L}^{(n\leftarrow n+m)}_d 
        e^{\mathbb{L}^{(n+m)}_0 \tau_1} \mathbb{L}^{(n+m\leftarrow n)}_u 
        e^{\mathbb{L}^{(n)}_0 \tau} | \rho^{(n)}_0 \rrangle 
        \right) + \mathcal{O}(\lambda^2).
    \end{split}
    \end{equation}
    \end{widetext}
    The leading-order contribution reduces to Eq.~\eqref{eq_lossjumpDQPT}, governed by the effective non-Hermitian Hamiltonian. 
    The first-order correction with respect to $\lambda$ consists of two parts. 
    The first term corresponds to a particle backflow process illustrated in Fig.~\ref{fig:1}: the initial state $|\rho_0^{(n)}\rrangle$ first evolves under $\mathbb{L}_0^{(n)}$ within the double Hilbert subspace $\mathcal{H}^{(n)}\otimes \mathcal{H}^{(n)}$, then transitions to the $\mathcal{H}^{(n-1)}\otimes \mathcal{H}^{(n-1)}$ subspace via the loss jump operator $\mathbb{L}_d^{(n-1\leftarrow n)}$, evolves under $\mathbb{L}_0^{(n-1)}$, and finally jumps back to the original subspace $\mathcal{H}^{(n)}\otimes \mathcal{H}^{(n)}$ through the gain process $\mathbb{L}_{u}^{(n\leftarrow n-1)}$. 
    The second term represents another particle backflow path similarly, as shown in Fig.~\ref{fig:1}. 
    Next, we prove that all many-body backflow contributions are strictly positive, derived from the completely positive and trace-preserving (CPTP)~\cite{GKS1976,Lindblad1976,Daniel2020Review} property of Liouvillian time evolution. 
    The positivity of backflow ensures $g(t)>0$ even when the zeroth-order term $\llangle \rho_0 | e^{\mathbb{L}^{(n)}_0 t} | \rho_0 \rrangle =0$.
    This means that DQPTs occur when only loss or gain is present, but are smeared out by particle backflow when both gain and loss coexist. 
    
    Define Liouvillian time-evolution map $\mathcal{E}_{t>0}(\rho):\rho(t_0) \rightarrow \rho(t_0+t)$. Since it is a CPTP map, according to Choi-Kraus theorem~\cite{Choi1975}, it can be written in Kraus form: $\mathcal{E}_{t>0}(\rho_0) = \sum_{\mu} K_{\mu,t} \rho_0 K_{\mu,t}^{\dagger}$. 
    Projecting onto $n$-particle sector: $\mathcal{E}_t^{(n)}(\rho_0) = \mathcal{P}_n \mathcal{E}_t(\rho_0) \mathcal{P}_n = \mathcal{P}_n \sum_{\mu} K_{\mu,t} \rho_0 K_{\mu,t}^\dagger \mathcal{P}_n = \sum_\mu (\mathcal{P}_n K_{\mu,t} \mathcal{P}_n) (\mathcal{P}_n \rho_0 \mathcal{P}_n) (\mathcal{P}_n K_{\mu,t}^\dagger \mathcal{P}_n) = \sum_{\mu} K_{\mu,t}^{(n)} \rho_0^{(n)} (K_{\mu,t}^{(n)})^{\dagger}$. 
    Therefore, $\mathcal{E}_t^{(n)}$ is a completely positive (CP) map.
    Additionally, both particle gain and loss jump superoperators are also CP maps since their actions are in Kraus form: $\sum_{\mu} L_{\mu}\rho L^{\dagger}_{\mu}$. 
    Consequently, the integrand in the time integrals in Eq.~\eqref{eq_backflow}—the particle backflow induced by gain and loss quantum jumps—corresponds to a matrix representation of a CP map $\sum_{\mu}\tr(\rho^\dagger_0 K_{\mu} \rho_0 K_{\mu}^\dagger)$. 
    The backflow terms are generally positive and thus make $g(t)$ in Eq.~\eqref{eq_backflow} positive even when the leading-order term $\llangle \rho_0 | e^{\mathbb{L}_0^{(n)} t} | \rho_0 \rrangle=0$. 
    Therefore, the presence of backflow will suppress the DQPT predicted by the effective non-Hermitian Hamiltonian. 
    This completes the proof of Theorem~\ref{thm_backflow}. 
\end{proof}

In general, particle backflow is strictly positive and suppresses DQPTs. We note that, in certain special cases, backflow can vanish even with both particle gain and loss. Here, we provide a more readily verifiable criterion for the absence of particle backflow: 
\begin{equation}\label{eq_backflowvanish}
    \begin{split}
    \mathbb{L}_u^{(n\leftarrow n-m)} \mathbb{L}_d^{(n-m\leftarrow n)} = \mathbb{L}_u^{(n\leftarrow n-m)} \mathbb{L}_0^{(n-m)} \mathbb{L}_d^{(n-m\leftarrow n)} = 0.
    \end{split}
\end{equation}
One example is the DQPTs in the Bogoliubov–de Gennes Hamiltonian under open quantum systems~\cite{FanHeng2018PRB,Sedlmayr2018PRB}. 
In such cases, particle loss induces both effective gain and loss processes for Bogoliubov quasiparticles, leading to particle backflow in our framework.
We note that, the specific setups considered in in Refs.~\cite{FanHeng2018PRB,Sedlmayr2018PRB} satisfy the backflow vanishing condition in Eq.~\eqref{eq_backflowvanish}, thereby enabling the occurrence of DQPTs. However, in general settings, particle loss in BdG Hamiltonians will smear out the DQPTs due to presence of the backflow effect. 

\emph{{\clr Large deviation behavior in long-time dynamics induced by backflow}.---}~We now apply the theorem to a two-band fermionic lattice model and show that an infinitesimal addition of particle gain not only smears out the DQPTs, but also induces a pronounced long-time deviation, a non-perturbative dynamical effect arising from many-body backflow. 

\begin{figure}[t]
    \centering
    \includegraphics[width=\linewidth]{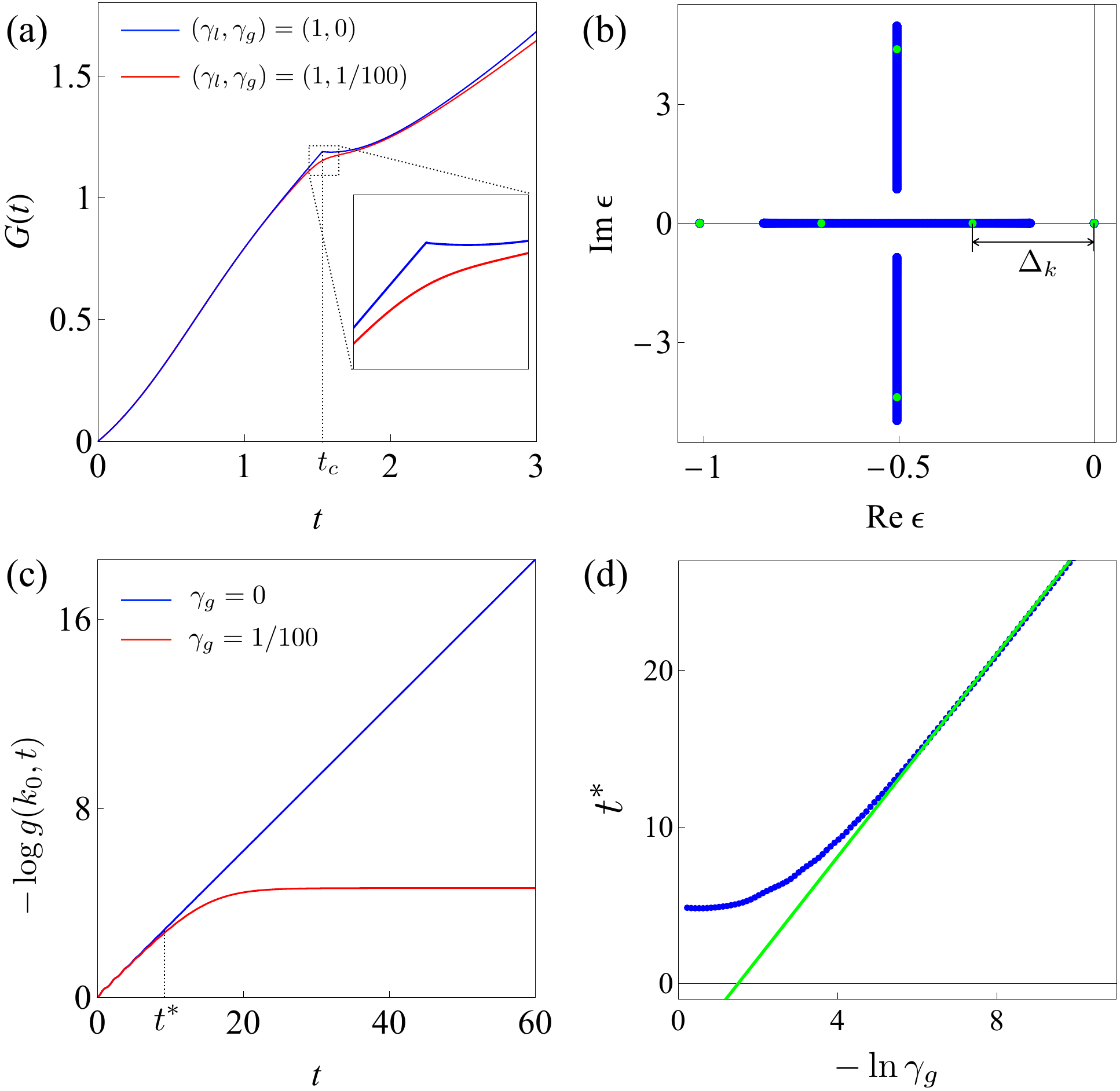}
    \caption{
    (a) The DQPT peak at $t=t_c$ is smeared out when a weak ($\gamma_g=1/100$) gain dissipation is introduced;
    (b) The blue dots show the Liouvillian spectrum for all $k$, and the green dots are the eigenvalues of the Liouvillian for a given $k_0=1$;
    (c) Long-time dynamical behavior of $-\ln g_{\lambda}(k,t)$ is abruptly changed after $t^*$ when $\gamma_g=1/100$ is introduced;
    (d) The crossover time $t^*$ scales linearly with $\ln \gamma_g$, where the green line is the linear fit with slope $1/\Delta_k$.
    }
    \label{fig:2}
\end{figure}

Consider a fermionic lattice model with the Hamiltonian 
\begin{equation}\label{eq_freeham}
    H = \sum_k \hat c^\dagger_{k} h(k) \hat c_{k} = \sum_k \hat c^\dagger_{k} (\textbf{d}(k) \cdot \bm{\sigma}) \hat c_{k}, 
\end{equation}
where $c^\dagger = (c^\dagger_{A}, c^\dagger_{B})$ is a two-component spinor of fermionic creation operator on $A$ and $B$ sublattices. 
The Hamiltonian $\textbf{d}(k) \cdot \bm{\sigma}$ has energy bands $\epsilon_{\pm}(k) = \pm |\textbf{d}(k)|$.
We take the initial state to be the ground state of the Hamiltonian with $\textbf{d}_0(k) = (t_0 + w \cos{k}, w \sin{k}, 0)$ and set $w=1$ and $t_0=1/2$. 
The initial density matrix is given by $\rho_0 = \otimes_k \rho_0(k) = \otimes_k|v_-(k)\rangle\langle v_-(k)|$, where $|v_-(k)\rangle$ denotes the lower-band Bloch state of initial Hamiltonian. 
At time $t=0$, the system begins to evolve under a Liouvillian superoperator characterized by the post-quench Hamiltonian $\textbf{d}_1(k) = (t_1+ w \cos{k}, w \sin{k}, 0)$ with $t_1=3/2$ and the particle-loss dissipator
\begin{equation}\label{eq_singleloss}
    L_x^l = \sqrt{\gamma_l} c_{x,A}.
\end{equation}
It's known that when the pre-quench and post-quench (effective non-Hermitian) Hamiltonians belong to distinct topological phases, DQPTs are guaranteed to occur. 
In open quantum systems, however, the effect of jump terms $\sum_x L_x^{l} \rho L_x^{l,\dagger}$ should be taken into account. 
According to Theorem~\ref{thm_backflow}, with only particle loss or gain, the DQPTs are fully captured by the effective non-Hermitian Hamiltonian. 
In the pure-loss case, we calculate $G(t)$ defined in Eq.~\eqref{eq_LoschmidtRate}. 
As indicated by the blue lines in Fig.~\ref{fig:2}(a), the cusps signal the occurrence of DQPTs at critical times $t_c$. 
In sharp contrast, when a weak gain dissipator is introduced,
\begin{equation}\label{eq_singlegain}
    L_x^g = \sqrt{\gamma_g} c^{\dagger}_{x,A},
\end{equation}
these non-analytic cusps are smeared out, as shown by the red lines in Fig.~\ref{fig:2}(a), despite the pre- and post-quench effective non-Hermitian Hamiltonians reside in different topological phases. 
This smearing effect is contributed from the particle backflow as illustrated in Fig.~\ref{fig:1} when both the particle gain and loss are present. 
The numerical simulations fully confirm Theorem~\ref{thm_backflow}. 
See details derivations in Sup.~Mat.~II~\cite{SupMat}.

We now introduce a novel dynamical phenomenon induced by particle backflow: adding even an infinitesimally small gain dissipator leads to a significant long-time deviation in $G(t)$, compared to the pure-loss case. 
Define $G_{\gamma_g}(t)$ be the Loschmidt rate with perturbative gain strength $\gamma_g$. The dynamical phenomenon can be expressed as: 
\begin{equation}
    \lim_{t\rightarrow\infty} \lim_{\gamma_g\rightarrow 0^+} G_{0}(t)-G_{\gamma_g}(t) \propto |\tilde{\Delta}| t,
\end{equation}
where $\tilde{\Delta}=\sum_k \Delta_k/N$ and $\Delta_k$ ($\Re{\Delta}_k<0$) denotes the Liouvillian gap at momentum $k$. 
The numerical result is shown in Fig.~\ref{fig:2}(b).
Next we explain the underlying reason of this non-perturbative dynamical phenomenon and leave the detailed derivations in Appendix~B. 

When only loss or gain is present, the initial density matrix $\rho_0$ must evolves to either a vacuum or fully occupied steady state, respectively.
Either steady state is orthogonal to the initial state, so the long-time dynamics of $G(t)$ is fully governed by the Liouvillian gap and has no steady-state contributions. 
However, when both gain and loss are present, the backflow effect modifies the steady state, distributing population across all particle-number Hilbert space. 
In this case, the long-time behavior is no longer dictated by the Liouvillian gap alone, but instead by the overlap between the initial and steady states. This is because the overlap between initial state and all other Liouvillian eigenstates than steady state decay exponentially due to the Liouvillian gap. 
This non-perturbative dynamical effect, induced by the backflow and contributed from steady state, is shown in Fig.~\ref{fig:2}(c). 
A crossover time $t^{\ast}$ marks the onset of a significant deviation in $G(t)$; for $t<t^{\ast}$, $G_0(t)$ and $G_{\gamma_g}(t)$ nearly coincide and the dynamics is dominated by the Liouvillian gap; while for $t>t^{\ast}$, the dynamics are dominated by the overlap between the initial and steady states, thus showing an exponential derivation in time from the case without gain. 
This crossover timescale is set by the competition between the initial–steady state overlap, proportional to $\gamma_g$, and the Liouvillian gap $\Delta_k$.
For a fixed momentum $k$, the crossover time is analytically estimated as (see derivations in Appendix~B)
\begin{equation}
    t_{\ast} \sim (\ln \gamma_g)/\Delta_k,
\end{equation}
which matches well with the numerical simulations in Fig.~\ref{fig:2}(d). 

Beyond the two-band model, we analytically solve the Hatsugai-Kohmoto model with spin-density interactions~\cite{hatsugai1992exactly} in Sup.~Mat.~III~\cite{SupMat}. 
We show that in the strong interaction limit, DQPTs can be exactly determined by analytically calculating the Fisher zeros---complex zeros of time $t$ in the Loschmidt rate function $G(t)$ given by Eq.~\eqref{eq_LoschmidtRate}. 
When both gain and loss are present, these DQPTs are smeared out, in agreement with our theorem. 

\begin{figure}[t]
    \centering
    \includegraphics[width=\linewidth]{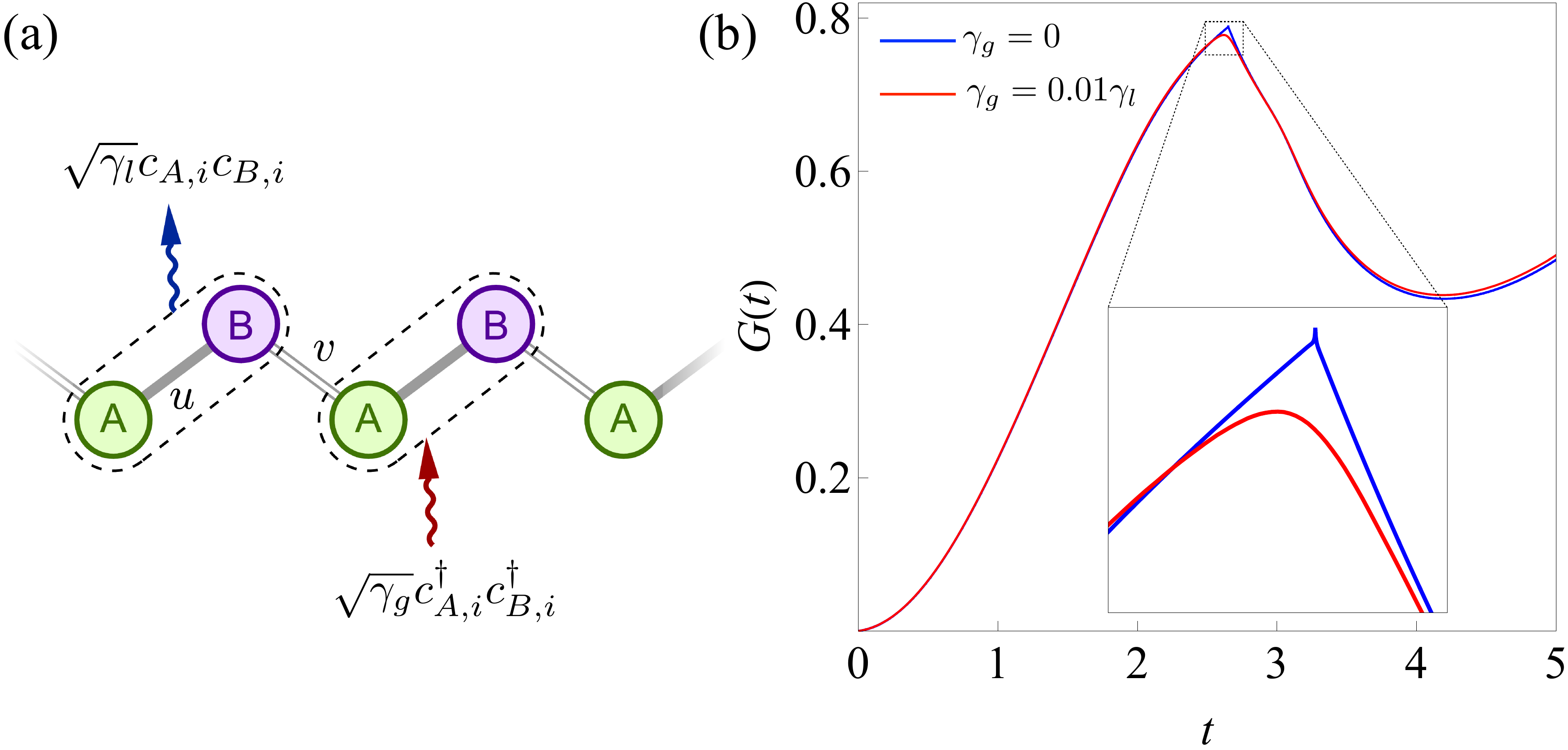}
    \caption{
    (a) Illustration of the model with two-body gain and loss.
    (b) The blue curve shows the flux-averaged DQPT without gain ($\gamma_l=0.4; \gamma_g=0$); The red curve shows that a tiny amount of gain ($\gamma_l=0.4; \gamma_g=\gamma_l/100$) smears out the many-body DQPT.
    Numerical setup: we calculate 7 unit cells for the 1D chain; In Eq.~\eqref{eq:fluxinserting}, the flux $\phi$ is sampled with interval $2\pi/750$; In Monte-Carlo wavefunction method, 1000 samples are used; the evolution time ranges from 0 to 5 with interval 0.005. 
    }
    \label{fig:3}
\end{figure}

\emph{{\clr A generic many-body model with two-body loss and gain}.---}~We now consider a quenched Liouvillian with general two-body gain and loss dissipators (Fig.~\ref{fig:3}(a)), given by
\begin{equation}
    \begin{split}
    \frac{d\rho}{dt} = \mathcal{L}\rho & = -i (H_{\text{eff}}\rho - \rho H_{\text{eff}}^{\dagger}) \\
    &+ \sum_x \gamma_{l} c_{x,A}c_{x,B} \rho c^{\dagger}_{x,B}c^{\dagger}_{x,A} + \gamma_{g} c^{\dagger}_{x,A}c^{\dagger}_{x,B} \rho c_{x,B}c_{x,A}
    \end{split}
\end{equation}
under periodic boundary conditions. 
The effective Hamiltonian is 
$H_{\text{eff}} = H - \frac{i}{2} (\gamma_l+\gamma_g) \sum_{x}n_{x,A}n_{x,B} - \frac{i}{2} \gamma_g \sum_{x} (1-n_{x,A}-n_{x,B})$, where $H=\sum_k c_{k}^{\dagger} \bm{d}_1(k)\cdot \bm{\sigma} c_{k}$ governs the unitary dynamics. 
Importantly, $H_{\text{eff}}$ respect symmetries: it is invariant under the transformation $c_{x,\alpha}\rightarrow c_{x,\alpha} e^{i\theta}$, $c_{x,\alpha}^{\dagger}\rightarrow c_{x,\alpha}^{\dagger} e^{-i\theta}$, and commutes with the total particle number operator $\hat N = \sum_{x,\alpha}n_{x,\alpha}$, i.e., $[H_{\text{eff}},\hat N]=0$. 
As a result, $\mathbb{L}_0 = - i (H_{\text{eff}} \otimes I - I \otimes H^{\ast}_{\text{eff}}) $ can be block-diagonalized into different particle number sectors $\mathbb{L}_0 = \bigoplus_N \mathbb{L}_0^{(N)}$. 
The two-body quantum jump operators change the particle number by two: the loss quantum jump reduces the particle number from $n$ to $n-2$, while the gain quantum jump process restores it back to the $n$-particle sector. The positivity of such particle backflow leads to a smoothing of DQPT signatures, effectively smearing out the non-analytic behavior.  
Now we verify the theorem numerically using Monte-Carlo wavefunction method (also known as the quantum trajectory method)~\cite{Molmer1993,Daley2014Review}. 

We first focus on the case with only particle loss ($\gamma_l\neq 0$ and $\gamma_g=0$). 
In this case, the jump operator does not contribute to the occurrence of DQPTs, instead, the dynamics are entirely governed by the effective non-Hermitian Hamiltonian. 
Due to the presence of the imaginary Hubbard-type interaction term $-\frac{i}{2}\gamma_l \sum_x n_{x,A}n_{x,B}$, the evolution is no longer decomposable into independent momentum sectors. 
To probe the DQPT, we insert a flux $\phi$ into the finite-size periodic chain and perform exact diagonalization of the many-body effective non-Hermitian Hamiltonian $H_{\text{eff}}(\phi)$. 
The initial state $|\Psi_0\rangle = \Pi_k b_{-,k}^{\dagger}|0\rangle$ is a Slater determinant, corresponding to the ground state of the pre-quench Hamiltonian $H = \sum_k c_{k}^{\dagger} \bm{d}_0(k)\cdot \bm{\sigma} c_{k}$, where $\hat b_{-,k}$ denotes the fermion creation operator in the lower energy band. 
The DQPT is characterized by the flux-averaged Loschmidt amplitude,
\begin{equation}\label{eq:fluxinserting}
    G(t) = \frac{1}{2\pi} \int_0^{2\pi} d\phi \ln \langle \Psi_0|e^{-i H_{\text{eff}}(\phi)t}|\Psi_0\rangle.
\end{equation}
The result is illustrated by the blue curve in Fig.~\ref{fig:3}(b), where a cusp indicates the occurrence of DQPT.  

Once particle gain is introduced, the effective non-Hermitian Hamiltonian alone is no longer sufficient to capture the system's dynamics. 
The full Liouvillian time evolution should be considered. 
To this end, we employ the Monte Carlo wavefunction method (see details of Methods in Sup.~Mat.~IV~\cite{SupMat}). 
Upon adding a small particle gain term with strength $\gamma_g=\gamma_l/100$, the cusp in the Loschmidt signal is washed out, as illustrated by the red curve in Fig.~\ref{fig:3}(b). 
This confirms that Theorem~\ref{thm_backflow} remains valid in a non-integrable many-body system with two-body gain and loss quantum jumps. 

\emph{{\clr Acknowledgment}.---}~This research was supported by the Office of Naval Research (Grant No. MURI N00014-20-1-2479) and the National Science Foundation through the Materials Research Science and Engineering Center at the University of Michigan (Award No. DMR-2309029). 

\appendix
\twocolumngrid

\section{End Matter}
\emph{{\clr Appendix A: Derivations of Many-Body Backflow in Eq.~(4).}---}~The time evolution operator can be expanded into the Dyson series
\begin{align}\label{eq_DysonSeries}
    e^{(A+\lambda B)t} & = e^{At} \mathcal{T}e^{\lambda \int_0^t d\tau B_I(\tau)} \nonumber \\
    & = e^{At} \left( 1 + \lambda\int_0^t d\tau \, e^{-A\tau} B e^{A\tau} \right. \nonumber \\ 
    & \left. + \lambda^2 \int_0^t d\tau_1\int_0^{\tau_1} d\tau_2 \, e^{-A\tau_1} B e^{A\tau_1} e^{-A\tau_2} B e^{A\tau_2} + \dots \right),
\end{align}
where $B_I(\tau) = e^{-A\tau}Be^{A\tau}$ and $\mathcal{T}$ refers to the time ordering operator. 
The Dyson series expansion can be derived by using the interaction picture of time evolution. 

We provide a short derivation in what follows. 
Define $U(t) = e^{(A+B)t} = e^{At}B(t)$. 
Then $B(t) = e^{-At}U(t)$. 
Taking the time derivative on both sides, we have 
\begin{equation}
\begin{split}
    \frac{d B(t)}{d t} & = - A B(t) + e^{-At} (A+B) e^{At}B(t)\\
    &= e^{-At} B e^{At}B(t) = B_I(t) B(t).
\end{split}
\end{equation}
Therefore, $B(t) = I + \int_0^t d\tau B_I(\tau) + \int_0^t d\tau_1 \int_{0}^{\tau_1} d\tau_2 B_I(\tau_1)B_I(\tau_2) + \dots$. 
Finally, we have $U(t)=e^{At}B(t) = e^{At}\mathcal{T}e^{\lambda \int_0^t dsB_I(s)}$. 

We now apply the Dyson series to Liouvillian time evolution in double Hilbert space. 
When the density matrix is vectorized by row-wise flattening, the Liouvillian superoperator becomes a linear operator with the matrix representation:
\begin{equation*}
    \begin{split}
    \mathbb{L} &= \mathbb{L}_0 + \mathbb{L}_d + \lambda \mathbb{L}_u \\
    & = - i (H_{\text{eff}} \otimes I - I \otimes H^{\ast}_{\text{eff}}) + \sum_{\mu} L^{l}_{\mu}\otimes L^{l,\ast}_{\mu} + \lambda \sum_{\mu} L^{g}_{\mu}\otimes L^{g,\ast}_{\mu},
    \end{split}
\end{equation*}
where $H_{\text{eff}} = H - \frac{i}{2}\sum_{\mu}L_\mu^\dagger L_\mu$, and $L^{l}_{\mu}, L^{g}_{\mu}$ are the matrix representations for $\mu$-th loss and gain jump dissipators, respectively. 
Suppose the effective non-Hermitian Hamiltonian has $U(1)$ symmetry, for example, total particle number conservation $[H_{\text{eff}}, N] = 0$ in fermion system or total magnetization conservation in spin system $[H_{\text{eff}}, S_z] = 0$. 
Therefore, $\mathbb{L}_0$ can be block diagonal, $\mathbb{L}_0 = \bigoplus_n \mathbb{L}_0^{(n)}$ and $n$ denotes each block given by the $U(1)$ symmetry charge. 
In this basis, the loss quantum jump process maps a density matrix in $n$-particle space to $m<n$-particle space, the matrix representation in double Hilbert space is $L^{l}_{\mu}\otimes L^{l,\ast}_{\mu}|\rho^{(n)}\rrangle \rightarrow |\rho^{(m)}\rrangle$. 

We treat $A = \mathbb{L}_0+\mathbb{L}_d$ and $B=\mathbb{L}_u$ in Eq.~\eqref{eq_DysonSeries}, and the full Liouvillian time evolution becomes
\begin{equation*}
    \begin{split}
    & U(t) = e^{(\mathbb{L}_0+\mathbb{L}_d+\lambda \mathbb{L}_u) t} = e^{(\mathbb{L}_0+\mathbb{L}_d)t} \mathcal{T} e^{\lambda \int_{0}^t d\tau e^{-(\mathbb{L}_0+\mathbb{L}_d)\tau} \mathbb{L}_u e^{(\mathbb{L}_0+\mathbb{L}_d)\tau}} \\
    & = e^{(\mathbb{L}_0+\mathbb{L}_d)t} + \lambda \, e^{(\mathbb{L}_0+\mathbb{L}_d)t} \int_{0}^t d\tau e^{-(\mathbb{L}_0+\mathbb{L}_d)\tau} \mathbb{L}_u e^{(\mathbb{L}_0+\mathbb{L}_d)\tau} + \mathcal{O}(\lambda^2) \\
    & = U_0(t) + \lambda \, \int_{0}^t d\tau U_0(t-\tau) \mathbb{L}_u U_0(\tau) + \mathcal{O}(\lambda^2)
    \end{split}
\end{equation*}
Now we use the Dyson series again for $U_0(t) = e^{(\mathbb{L}_0+\mathbb{L}_d)t}$, it becomes
\begin{equation*}
    U_0(t) = e^{(\mathbb{L}_0+\mathbb{L}_d)t} = e^{\mathbb{L}_0 t} \left( 1 + \int_0^t  d\tau e^{-\mathbb{L}_0 \tau} \mathbb{L}_d e^{\mathbb{L}_0 \tau} + \mathcal{O}(\mathbb{L}_d^2) \right).
\end{equation*}
DQPT is characterized by the quantity $\llangle\rho_0|U(t)|\rho_0\rrangle$, and the initial density matrix state is supposed to have a fixed particle number $n_0$. 
Therefore, for the zeroth-order term of $\lambda$ in $U(t)$---namely $U_0(t)$---all terms involving $\mathbb{L}_d$ in $U_0(t)$ have no contribution to this quantity, and we have $\llangle\rho_0|U_0(t)|\rho_0\rrangle = \llangle\rho_0|e^{\mathbb{L}^{(n_0)}_0 t}|\rho_0\rrangle$, which is Eq.~\eqref{eq_lossjumpDQPT} in the main text. 
For the first-order term of $\lambda$ in $U(t)$, there are two terms make $\llangle\rho_0|U(t)|\rho_0\rrangle$ nonzero. 
First, preserve $U_0(t-\tau)$ to first-order of $\mathbb{L}_d$ and $U_0(\tau)$ to zeroth-order term. 
Second, preserve $U_0(t-\tau)$ to zeroth-order term and $U_0(\tau)$ to first-order term in $\mathbb{L}_d$. 
All other terms have no contribution to $\llangle\rho_0|U(t)|\rho_0\rrangle$. 
These two terms correspond to Eq.~\eqref{eq_backflow} in the main text, which physically corresponds to the partial backflow process in the double Hilbert space as illustrated in Fig.~\ref{fig:1}. 

For more general jump dissipators, such as $\hat L^l = \sum_{s} \sqrt{\gamma_s^l}\hat c_1\dots \hat c_s$ and $\hat L^g = \sum_{s} \sqrt{\gamma_s^g}\hat c_1^{\dagger}\dots \hat c_s^{\dagger}$, the first-order contribution in $\lambda$ to $U(t)$ involves the $\mathbb{L}_u^{(m\leftarrow n)}$ and $\mathbb{L}_d^{(p\leftarrow q)}$ blocks, which can collectively construct the particle backflow in the double Hilbert space. Theorem~\ref{thm_backflow} remains valid in this general setting. 

\emph{{\clr Appendix B: Deviations of Eqs.~(9) and (10)}---}~Here, we provide more details on the large deviation in long-time dynamics of $G(t)$ when the particle gain is perturbatively introduced.  

At the long-time limit, we observe the following scaling behavior:
\begin{equation}
    \lim_{t\rightarrow\infty} \lim_{\lambda\rightarrow 0^+} G_{0}(t)-G_{\lambda}(t) \propto |\tilde{\Delta}| t,
\end{equation}
where $\tilde{\Delta}$ denotes the averaged Liouvillian gap density, defined as $$\tilde{\Delta} = \frac{\sum_k \Delta_k}{N},$$ with $\Delta_k$ representing the Liouvillian gap at momentum $k$, and $N$ the total number of degrees of freedom. 
Here, $\lambda$ denotes the strength of particle gain, which is $\gamma_g$ in Eq.~\eqref{eq_singlegain}. 
For small $\lambda \ll 1$, the Liouvillian gap remains nearly unchanged. Thus, the significant deviation in long-time dynamics is not due to a change in the Liouvillian gap, but rather results from the modification of the steady state induced by particle backflow. 

Now we prove this long-time limit behavior. 
By definition, the Loschmidt rate function in Eq.~\eqref{eq_LoschmidtRate} can be expressed as
\begin{equation}
    \begin{split}
    G_{\lambda}(t) & \equiv \lim_{N\rightarrow \infty} -\frac{1}{N} \sum_k \ln g_{\lambda}(k,t) \\ 
    & = \lim_{N\rightarrow \infty} -\frac{1}{N} \sum_{k} \, \ln \llangle \rho_0(k) | e^{\mathbb{L}_k(\lambda) t}|\rho_0(k) \rrangle,
    \end{split}
\end{equation}
where $\mathbb{L}_k(\lambda) = \mathbb{L}_0(\lambda) + \mathbb{L}_d + \lambda \mathbb{L}_u$, and $\lambda$ is the strength of the particle gain. 
Since $\mathbb{L}_k(\lambda)$ is a non-Hermitian matrix in the double Hilbert space, it can be decomposed using the biorthogonal basis, namely, $\mathbb{L}_k(\lambda) = \sum_n \epsilon_n(k,\lambda) |\psi_n^R(k)\rrangle\llangle\psi_n^L(k)|$. 
Therefore, we have $\llangle \rho_0(k)|e^{\mathbb{L}_k(\lambda) t}|\rho_0(k)\rrangle = \sum_{n} e^{\epsilon_n(k,\lambda)t} \llangle\rho_0(k)|\psi_n^R(k)\rrangle\llangle\psi_n^L(k)|\rho_0(k)\rrangle$. 
It's worth noting that, when $\lambda=0$, there is no particle gain, the steady state is the vacuum state. 
In this case, the initial density matrix has no overlap with the steady state. 
Due to the absence of overlap with the steady state, the return function can be expanded as:  
\begin{equation}
    g_{0}(k,t) = \llangle \rho_0(k)|e^{\mathbb{L}_k(\lambda=0) t}|\rho_0(k)\rrangle = e^{\Delta_k t} C_1(k) + \sum_{n=2}^M e^{\epsilon_n(k) t} C_n(k),
\end{equation}
where $C_n = \llangle\rho_0|\psi_n^R\rrangle\llangle\psi_n^L|\rho_0\rrangle$ and the Liouvillian eigenvalues have been ordered according to their real parts: $\epsilon_0=0 \geq \epsilon_1\equiv\Delta_k \geq \epsilon_2 \geq \dots \epsilon_M$. 
In general, a finite Liouvillian gap $\Re\Delta_k<0$ is assumed. 
When we turn on the particle gain with strength $\lambda$, the following points should be carefully considered:
(i.)~The initial state has a finite overlap with steady state, and the amplitude of the overlap scales with $\lambda$;  
(ii.)~For infinitesimally small gain ($\lambda \ll 1$), the Liouvillian gap is only weakly perturbed, justifying a perturbative treatment;
(iii.)~Contributions from other Liouvillian eigenstates with larger real parts can be neglected, as we retain only the leading-order term that dominates the long-time Liouvillian dynamics. 
For these reasons, we can expand the return function with nonzero $\lambda$ as follows:
\begin{equation}
    g_{\lambda}(k,t) =\llangle \rho_0(k)|e^{\mathcal{L}_k(\lambda) t}|\rho_0(k)\rrangle \sim \lambda^2 C_0(k) + C_1(k) e^{\Delta_k t}e^{\lambda \Delta^{\prime}_k(0)t} + \dots
\end{equation}
Here, for a small $\lambda$, we expand the Liouvillian gap as $\Delta_k(\lambda) = \Delta_k(\lambda=0) + \lambda \partial_{\lambda}\Delta_k(\lambda)|_{\lambda=0} + \dots \approx \Delta_k$ and keep the first-order term for a finite Liouvillian gap $\Delta_k$. 
Finally, the difference between the cases with and without particle gain becomes
\begin{equation}
\begin{split}
    G_0(t) - G_{\lambda}(t) 
    & = \lim_{N\rightarrow \infty} -\frac{1}{N} \sum_k \ln \frac{g_{0}(k,t)}{g_{\lambda}(k,t)} \\
    & \sim \lim_{N\rightarrow \infty} -\frac{1}{N} \sum_k \ln[\frac{C_1(k)e^{\Delta_k t}}{\lambda^2 C_0(k) + C_1(k) e^{\Delta_k t} }] \\
    & = \lim_{N\rightarrow \infty} -\frac{1}{N} \sum_k \ln[\frac{1}{\lambda^2 B_0(k) e^{-\Delta_k t} + 1}] \\ 
    & = \lim_{N\rightarrow \infty} \frac{1}{N} \sum_k \ln[\lambda^2 B_0(k) e^{-\Delta_k t} + 1],
\end{split}
\end{equation}
where $B_0(k) = C_0(k)/C_1(k)$ and $\Re\Delta_k < 0$. 
From the expression, we arrive at the following conclusions:
(i.)~When $\lambda$ is exactly zero, $G_0(t) - G_{\lambda=0}(t) = 0$;
(ii.)~For a given infinitesimally small $\lambda$, at long-time limit $t \rightarrow +\infty$, $e^{-\Delta_k t}$ dominates dynamics and it becomes  
$$\lim_{t\rightarrow \infty} G_0(t) - G_{\lambda}(t) \sim \lim_{N\rightarrow \infty} \frac{1}{N} \sum_k (-\Delta_k t ) = |\tilde{\Delta}| t.$$ 
The crossover time between perturbative and non-perturbative regime is characterized by the time when $\lambda^2 e^{-\Delta_k t}$ cannot be treated as perturbation. 
Therefore, the crossover time can be estimated as $\lambda^2 e^{-\Delta_k t_c}\sim 1$, and thus 
\begin{equation}
    t_c \sim \frac{\ln \lambda}{\Delta_k}.
\end{equation}


%


\appendix
\onecolumngrid
\newpage
\begin{center}
    \textbf{Supplemental Material}
\end{center}

\section{I.~~Quantum dynamical phase transitions in the thermodynamic limit}\label{SecI}

In this section, we analyze the nonanalytic behavior of the Loschmidt rate function defined in Eq.~(2) of the main text, and explain why it can be viewed as an analog of an equilibrium phase transition in the thermodynamic limit. 

The Loschmidt rate function is defined as
\begin{equation}
    \begin{split}
    G(t) &= \lim_{N\rightarrow +\infty} -\frac{1}{N} \sum_k \ln g(k,t) \\ 
    &= \lim_{N\rightarrow +\infty} -\frac{1}{N} \sum_k \ln \tr[\rho_0(k)\rho(k,t)].
    \end{split}
\end{equation}
In the thermodynamic limit, the momentum summation becomes an integral
\begin{equation}\label{eq_LoschmidtLimit}
    G(t) = -\frac{1}{2\pi} \int dk \ln g(k,t).
\end{equation}
Since $g(k,t)$ represents the overlap between two density matrices, it is typically an analytic function of both $k$ and $t$. 
However, nonanalytic behavior arises when $g(k,t)=0$, which causes a logarithmic divergence. 
Assuming $g(k,t)=0$ at some critical point $(k_c,t_c)$, we expand $g(k,t)$ near this point by setting $q=k-k_c$ and $\tau=t-t_c$, yielding the general form
\begin{equation}\label{eq_taylorphase}
    g(q,\tau) \sim \sum_{m,n} q^{2m} + B_n \tau^{2n}.
\end{equation}
Odd powers are excluded due to the local minimum at $(k_c,t_c)$ and the positivity $0\leq g(k,t)\geq 1$ from its probabilistic meaning. 

Without loss of generality, we consider only the leading terms with $m=n=1$, and set $B_1=1$, since the prefactor does not influence the nonanalytic structure.
Then, near the critical point,
\begin{equation}
    g(q,\tau) \sim q^{2} + \tau^{2}.
\end{equation}
At $q=0$ ($k=k_c$), the logarithmic term $\ln g(q,\tau)$ diverges as $\tau\rightarrow 0$.
Integrating over momentum yields the Loschmidt rate function
\begin{equation}
    G(\tau) = -\frac{1}{2\pi} \int_{-\Delta}^{\Delta} dq \ln (q^2 + \tau^2) = -\frac{4 \Delta - 4 \tau \arctan{\frac{\Delta}{\tau}} -2 \Delta \ln (\Delta^2+\tau^2)}{2\pi},
\end{equation}
where $\Delta$ is a finite cutoff. 
It follows that
\begin{equation}
    \begin{split}
    & G(\tau = 0) = \frac{2\Delta}{\pi} (\ln{\Delta} -1); \\
    & \lim_{\tau\rightarrow 0^{\mp}}\frac{d G(\tau)}{d\tau} = \pm 1.
    \end{split}
\end{equation}
Thus, the rate function is continuous at $t_c$, but its first derivative with respect with $t$ is discontinuous. 

Interpreting $G(t)$ as the dynamical analog of a boundary partition function in equilibrium systems, this nonanalyticity therefore signifies a dynamical quantum phase transition (DQPT). 
Although we focused on the minimal case $m=n=1$, the same conclusion holds for more general cases in Eq.~\eqref{eq_taylorphase}, which we do not elaborate here. 
This minimal model illustrates the emergence of nonanalyticity near DQPTs, highlighting the necessity of defining the Loschmidt rate function in the thermodynamic limit.
In the same spirit, for the non-integrable model discussed in the main text, we employ flux averaging to mimic the thermodynamic limit [Eq.~(12) in the main text], allowing similar nonanalytic features to emerge near the critical time.

\section{II.~~Full Liouvillian representation in double occupation basis}\label{SecII}
In this section, we demonstrate how to represent the full Liouvillian in the double occupation basis using the two-band example in the main text. 
The density matrices live in the operator space with basis $\{\ket{\Psi_n}\bra{\Psi_{n'}}\}$ where $\Psi_n$ is a many-body eigenstate in the Fock space.
We flatten this basis into $\{\ket{\Psi_n}\otimes\ket{\Psi_{n'}}\}$ or simply denoted as $\{\ket{\Psi_n;\Psi_{n'}}\}$ in the double occupation basis, leading to a vectorized representation of the density matrices.
In this representation, the Liouvillian superoperator becomes a matrix with elements given by $\mathbb{L}_{nn',mm'}=\bra{\Psi_n;\Psi_{n'}}\mathcal{L}\ket{\Psi_m;\Psi_{m'}}$.
To construct the Liouvillian operator, we could use the third quantization formalism~\cite{Tomaz2008}, which allows us to express the Liouvillian in terms of the third quantized Hamiltonian and jump operators.
However, to construct the matrix representation for the Liouvillian in the momentum space, it comes handy to directly use the following identity:
\begin{equation}\label{eq:vec_identity}
    \operatorname{vec}(A\rho B)= (A\otimes B^T)\operatorname{vec}(\rho). 
\end{equation}
where we choose the convention that $\rho=(\rho_{ij})$ is flattened as $\operatorname{vec}(\rho)=(\rho_{11},\rho_{12},\cdots,\rho_{1N},\cdots,\rho_{NN})$.
(In some other reference, the vectorization could be defined as $\operatorname{vec}(\rho)=(\rho_{11},\rho_{21},\cdots,\rho_{N1},\cdots,\rho_{NN})$, which leads to $\operatorname{vec}(A\rho B)=(B^T\otimes A)\operatorname{vec}(\rho)$, but we stick to the former convention here due to its compatiblity with the \texttt{Flatten[]} function in Mathematica.)
Using this identity, we can write the Liouvillian superoperator in the double occupation basis as
\begin{equation}
    \mathbb{L}=-i(H_{\text{eff}}\otimes I-I\otimes H_{\text{eff}}^{\ast}) + \sum_{\mu} L_\mu\otimes L_\mu^*.
\end{equation}
Next we show a concrete example of the Liouvillian representation in the double occupation basis for the two-band model in the main text.
In this case, the initial state is a direct product of all momentum sectors for the lower band, i.e., $\rho_0(k)=\ket{v_{-}(k)}\bra{v_{-}(k)}$, where $|v_{-}(k)\rangle$ is the lower band eigenstate of the single-particle Hamiltonian $h(k)=\mathbf{d}(k)\cdot \boldsymbol{\sigma}$.
The Fock space for each $k$ momentum sector is spanned by the four states: $\mathcal{F}_k=\{\ket{0},c^\dagger_{A,k}\ket{0},c^\dagger_{B,k}\ket{0},c_{A,k}^\dagger c^\dagger_{B,k}\ket{0}\}$. 
We could simply label these states as $\{\ket{00},\ket{10},\ket{01},\ket{11}\}$.
\begin{equation}
\underbrace{\ket{00}}_{\mathcal{H}^{(0)}},\underbrace{\ket{10},\ket{01}}_{\mathcal{H}^{(1)}},\underbrace{\ket{11}}_{\mathcal{H}^{(2)}}
\end{equation}
where $\mathcal{H}^{(n)}$ is the $n$-particle Hilbert space spanned by the $n$-particle states.
Then the double occupation space is spanned by $(\dim \mathcal{F}_k)^2=16$ states, labeled by $\ket{n_A,n_B;\overline{n}_A,\overline{n}_B}$.
However, due to the weak symmetry $\mathcal{U}_w(\rho) = e^{i\phi\hat{N}}\rho e^{-i\phi \hat{N}}$ that commutes with the Liouvillian, the subspaces with a fixed particle number difference $n_{\text{diff}}=n_A+n_B-\overline{n}_A-\overline{n}_B$ are invariant under the Liouvillian dynamics.
Hence, we can further restrict ourselves to the subspace with $n_{\text{diff}}=0$, which is spanned by the following 6 states:
\begin{equation}
\underbrace{\ket{00;00}}_{\mathcal{H}^{(0)}\otimes\mathcal{H}^{(0)}},\underbrace{\ket{10;10},\ket{10;01},\ket{01;10},\ket{01;01}}_{\mathcal{H}^{(1)}\otimes\mathcal{H}^{(1)}},\underbrace{\ket{11;11}}_{\mathcal{H}^{(2)}\otimes\mathcal{H}^{(2)}}.
\end{equation}
We are now ready to construct the Liouvillian superoperator in this double occupation basis.
The effective non-Hermitian Hamiltonians for each particle number sector are given by
\begin{equation}
    H_{\text{eff}}^{(0)}(k) = -\frac{i\gamma_g}{2},\ H_{\text{eff}}^{(1)}(k) = \mqty(-\frac{i\gamma_l}{2}&d_x(k)-i d_y(k)\\d_x(k)+i d_y(k)&-\frac{i\gamma_g}{2}),\ H_{\text{eff}}^{(2)}(k) = -\frac{i\gamma_l}{2}.
\end{equation}
The jump operators are given by
\begin{equation}
    L^{(0\leftarrow 1)}(k) = \mqty(\sqrt{\gamma_l}& 0),\ L^{(1\leftarrow 0)}(k) = \mqty(\sqrt{\gamma_g}\\0),\ L^{(1\leftarrow 2)}(k) = \mqty(0\\ \sqrt{\gamma_l}),\ L^{(2\leftarrow 1)}(k) = \mqty(0 &\sqrt{\gamma_g}),
\end{equation}
where the superscript $({n \leftarrow m})$ denotes the jump operator that shifts states from $m$-particle to $n$-particle sector. 
According to the identity in Eq.~\eqref{eq:vec_identity}, we have:
\begin{eqnarray}
    \mathbb{L}_0^{(0)} &=& -i(H_{\text{eff}}^{(0)}\otimes I-I\otimes H_{\text{eff}}^{(0)*})=-\gamma_g,\\
    \mathbb{L}_0^{(1)} &=& -i(H_{\text{eff}}^{(1)}\otimes I-I\otimes H_{\text{eff}}^{(1)*})=\mqty(-\gamma_l&i d_x+ d_y&-i d_x-d_y&0\\ i d_x-d_y&-\frac{1}{2}(\gamma_g+\gamma_l)&0&-i d_x-d_y\\-i d_x+d_y&0&-\frac{1}{2}(\gamma_g+\gamma_l)& i d_x+d_y\\0&-i d_x+d_y&i d_x-d_y&-\gamma_g),\\
    \mathbb{L}_0^{(2)} &=& -i(H_{\text{eff}}^{(2)}\otimes I-I\otimes H_{\text{eff}}^{(2)*})=-\gamma_l,\\
\end{eqnarray}
\begin{eqnarray}
    \mathbb{L}^{(0\leftarrow 1)} &=& L^{(0\leftarrow 1)}\otimes L^{(0\leftarrow 1)*}=\mqty(\gamma_l&0&0&0),\\
    \mathbb{L}^{(1\leftarrow 0)} &=& L^{(1\leftarrow 0)}\otimes L^{(1 \leftarrow 0)*}=\mqty(\gamma_g&0&0&0)^T,\\
    \mathbb{L}^{(1\leftarrow 2)} &=& L^{(1\leftarrow 2)}\otimes L^{(1\leftarrow 2)*}=\mqty(0&0&0&\gamma_l)^T,\\
    \mathbb{L}^{(2\leftarrow 1)} &=& L^{(2\leftarrow 1)}\otimes L^{(2\leftarrow 1)*}=\mqty(0&0&0&\gamma_g).
\end{eqnarray}
Assembling the Liouvillian superoperator in the double occupation basis, we have
\begin{equation}
    \mathbb{L} = \mqty(\mathbb{L}_0^{(0)} & \mathbb{L}^{(0\leftarrow 1)} & \mathbb{L}^{(0\leftarrow 2)} \\ \mathbb{L}^{(1\leftarrow 0)} & \mathbb{L}_0^{(1)} & \mathbb{L}^{(1\leftarrow 2)} \\ \mathbb{L}^{(2\leftarrow 0)} & \mathbb{L}^{(2\leftarrow 1)} & \mathbb{L}_0^{(2)})=\mqty(-\gamma_g&\gamma_l&0&0&0&0\\\gamma_g&-\gamma_l&id_x+d_y&-id_x-d_y&0&0\\0&i d_x-d_y&-\frac{1}{2}(\gamma_g+\gamma_l)&0&-i d_x-d_y&0\\0&-id_x+d_y&0&-\frac{1}{2}(\gamma_g+\gamma_l)&id_x+d_y&0\\0&0&-id_x+d_y&id_x-d_y&-\gamma_g&\gamma_l\\0&0&0&0&\gamma_g&-\gamma_l).
\end{equation}

\section{III.~~Fisher zeros in Hatsugai-Kohmoto interacting model}\label{SecIII}

To test the validity of the theorem for interacting systems, we introduce a spin-density interaction to the free Hamiltonian
\begin{equation}
    H_0 = \sum_k \hat c^\dagger_{k} h(k) \hat c_{k} = \sum_k \hat c^\dagger_{k} \begin{pmatrix}
    \textbf{d}(k) \cdot \bm{\sigma} & 0 \\ 0 & \textbf{d}(k) \cdot \bm{\sigma}
    \end{pmatrix} \hat c_{k}, 
\end{equation}
where $c^\dagger = (c^\dagger_{A,\uparrow}, c^\dagger_{B,\uparrow},c^\dagger_{A,\downarrow}, c^\dagger_{B,\downarrow})$ is a four-component spinor of fermionic creation operator. 
The term $\textbf{d}(k) \cdot \bm{\sigma}$ describes the single-spin Hamiltonian, with energy bands $\epsilon_{\pm}(k) = \pm |\textbf{d}(k)|$. 
The resulting Hamiltonian takes the form
\begin{equation}
    H_{\text{HK}}(k) = \sum_{k,\sigma=\uparrow\downarrow,\alpha=\pm} \epsilon_{\alpha}(k) \hat n_{\alpha,k,\sigma} + U (\hat n_{+,k,\leftarrow} + \hat n_{-,k,\leftarrow})(\hat n_{+,k,\rightarrow} + \hat n_{-,k,\rightarrow}),
\end{equation}
where $\epsilon_{\alpha = \pm}$ denote the free-fermion energy bands, and $c^{\dagger}_{\leftarrow} = \frac{1}{\sqrt{2}}(c^{\dagger}_{\uparrow} - c^{\dagger}_{\downarrow})$, $c^{\dagger}_{\rightarrow} = \frac{1}{\sqrt{2}}(c^{\dagger}_{\uparrow} + c^{\dagger}_{\downarrow})$. 
Note that the Hamiltonian $H_{\text{HK}}(k)$ respects momentum conservation. 
This type of interaction was introduced in Ref.~\cite{hatsugai1992exactly} to model long-range interactions in real space and admits fully analytical solutions. 

Assume that the pre-quench Hamiltonian is given by $H_{\text{HK}}(k)$ with $\textbf{d}_0(k) = (1/2 + \cos{k},0,\sin{k})$ and $U=0$, and the system is initially prepared in its half-filled ground state. 
Due to translation symmetry, the dynamics decouple in momentum space, allowing each momentum sector $k$ to evolve independently. 
We therefore work in momentum space and represent the four-band Hamiltonian in the occupation number basis for each $k$. 
The basis states are labeled as $|n_{A,\uparrow}n_{B,\uparrow}n_{A,\downarrow}n_{B,\downarrow}\rangle$. 
At half filling, the Fock space for each $k$ is spanned by the 6 two-particle states $\{|1100\rangle,|1010\rangle,|1001\rangle,|0110\rangle,|0101\rangle,|0011\rangle\}$. 
Let $h_0(k) = \textbf{d}_0(k)\cdot \bm{\sigma}$ be the pre-quench single-particle Hamiltonian. 
It has two spin-degenerate bands, with eigenstates $|v_{\pm}(k)\rangle$ satisfying $h_0(k)|v_{\pm}(k)\rangle = \epsilon_{0,\pm}(k)|v_{\pm}(k)\rangle$, and $\epsilon_{0,\pm}(k) = \pm|\textbf{d}_0(k)|$. 
The half-filling ground state is a direct product state: $|\Psi_0\rangle = \Pi_{k,\sigma} \hat b^{\dagger}_{k,-,\sigma}|0\rangle$, where $\hat b^{\dagger}_{k,-,\sigma}$ creates a fermion with spin $\sigma=\uparrow,\downarrow$ in the lower band. 
The band operator is related to the site basis by the basis transformation $\hat b^{\dagger}_{k,-,\sigma} = v_{-}^A c^{\dagger}_{k,A,\sigma} + v_{-}^B c^{\dagger}_{k,B,\sigma}$, where $v_{-}^{A/B}$ are the components of the Bloch wavefunction $|v_{-}(k)\rangle$. 
In the two-particle occupation basis defined above, the initial state takes the form
\begin{equation}\label{eq_initialstate}
    \Psi_0 = \{0,(v_{-}^A)^2, v_{-}^Av_{-}^B, v_{-}^A v_{-}^B, (v_{-}^B)^2, 0\}.
\end{equation}

\begin{figure*}[t]
    \begin{centering}
    \includegraphics[width=1\linewidth]{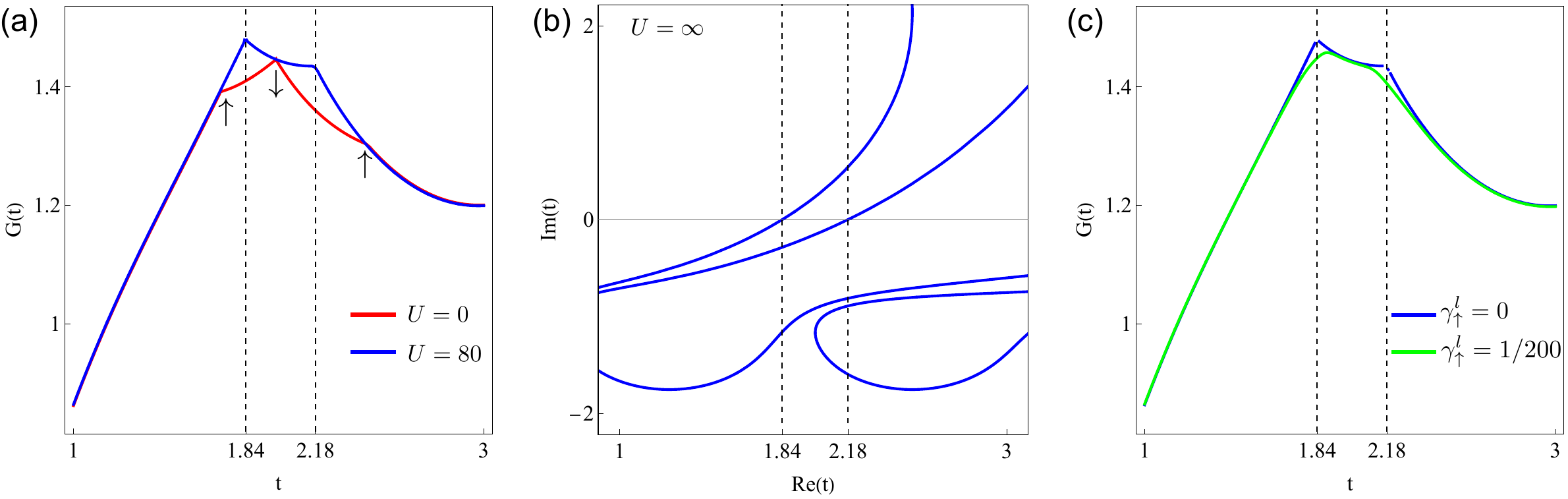}
    \par\end{centering}
    \protect\caption{\label{fig:S1} DQPTs in the Hatsugai-Kohmoto Interacting Model. 
    (a)~DQPTs for the interaction-free (red line) and Hatsugai-Kohmoto interacting (blue line) post-quench Hamiltonians.
    In the interaction-free case, spin degrees of freedom are decoupled, and each cusp can be labeled by either spin-$\uparrow$ or spin-$\downarrow$. 
    For the interacting case, a large interaction strength $U=80$ is used to approximate the infinite interaction limit. 
    (b)~Fisher zeros in the infinite-interaction limit. The intersections between the Fisher-zero algebraic curves and the real-time axis correspond to the critical times at which DQPTs occur in (a). 
    (c)~The Loschmidt rate function $G(t)$ for $(\gamma_{\uparrow}^g,\gamma_{\uparrow}^l) = (1/2, 0)$ (blue line) and $(\gamma_{\uparrow}^g,\gamma_{\uparrow}^l) = (1/2, 1/200)$ (green line). 
    The addition of a small particle-loss term clearly smears out the DQPT cusps.}
\end{figure*}

We first consider a quench process in which the initial state $|\Psi_0\rangle$ evolves under an open quantum system with the single-particle Hamiltonian $\textbf{d}_1(k) = (3/2 + \cos{k},0,\sin{k})$ and a spin-dependent gain jump operator $L_{x}^g = \sqrt{\gamma^g_{\uparrow}}\hat{c}^{\dagger}_{x,A,\uparrow}$. 
Without spin density interaction ($U=0$), the Liouvillian superoperator preserves $S_z$ symmetry, allowing the full dynamics to decouple into independent spin sectors. 
In the spin-$\downarrow$ sector, the quench process reduces to a unitary evolution: $h_{0,\downarrow}(k)=\textbf{d}_0\cdot \bm{\sigma} \rightarrow h_{1,\downarrow}(k)=\textbf{d}_1\cdot \bm{\sigma}$. 
In contrast, the spin-$\uparrow$ sector experiences both coherent and dissipative dynamics, effectively described by: $h_{0,\uparrow}(k)=\textbf{d}_0\cdot \bm{\sigma} \rightarrow h_{1,\uparrow}(k)=\textbf{d}_1\cdot \bm{\sigma}-\frac{i}{2}\gamma^g_{\uparrow}(\sigma_0 - \sigma_z)/2$. 
We compute the Loschmidt rate function $G(t)$ for both spin components, as shown in Fig.~\ref{fig:S1}(a). 

We now consider a second quench process, in which the initial state $|\Psi_0\rangle$ evolves under the Liouvillian superoperator characterized by $\textbf{d}_1(k) = (3/2 + \cos{k},0,\sin{k})$, spin-dependent gain dissipator $L_{x}^g = \sqrt{\gamma^g_{\uparrow}}\hat{c}^\dagger_{x,A,\uparrow}$, and a infinite repulsive interaction strength $U=+\infty$. 
The spin density interaction mixes spins, and we consider the effective two-particle Hamiltonian in the occupation basis. 
Since the initial state in Eq.~\eqref{eq_initialstate} has no components on the basis $|1100\rangle$ and $|0011\rangle$, we simplify the two-particle basis into four-dimensional, composed of $\{|1010\rangle,|1001\rangle,|0110\rangle,|0101\rangle\}$.
Under this basis, the post-quench two-particle effective non-Hermitian Hamiltonian can be written as 
\begin{equation}\label{eq_2bodyham}
    H_{\text{eff}}^{(2)} = 
    \begin{pmatrix}
    4V + 2d_{0,z} & d_{0,x} & d_{0,x} & 0 \\
    d_{0,x} & 2V & 2V & d_{0,x} \\
    d_{0,x} & 2V & 2V - \frac{i}{2}\gamma^g_{\uparrow} & d_{0,x} \\
    0 & d_{0,x} & d_{0,x} & 4V - 2 d_{0,z} - \frac{i}{2}\gamma^g_{\uparrow}
    \end{pmatrix}.
\end{equation}
Theorem~1 in the main text asserts that when only loss or gain dissipators are present, DQPTs are fully characterized by the effective non-Hermitian Hamiltonian: $g(k,t) = \tr[\rho_0 e^{-iH_{\text{eff}}t}\rho_0 e^{iH^{\dagger}_{\text{eff}}t}]$ and $\rho_0 = |\Psi_0\rangle\langle \Psi_0|$. 
In the limit of infinite interaction strength, the effective Hamiltonian in Eq.~\eqref{eq_2bodyham} decouples into singlet and triplet subspaces.
The singlet state is given by the antisymmetric combination $\frac{1}{\sqrt{2}}(|1001\rangle - |0110\rangle)$, which can be written as $\frac{1}{\sqrt{2}}(c_{A\uparrow}^\dagger c_{B\downarrow}^\dagger - c_{B\uparrow}^\dagger c_{A\downarrow}^\dagger)|0\rangle$, with eigenenergy $\epsilon_0=-\gamma^g_{\uparrow}/4$. 
The remaining three states form the triplet subspace. Since the singlet state is antisymmetric under spin exchange, it is orthogonal to the initial state, which is symmetric.
As a result, the singlet state does not contribute to the Loschmidt amplitude and thus plays no role in the DQPTs.
We therefore project the effective two-particle Hamiltonian $H^{(2)}_{\text{eff}}$ into the triplet subspace via
\begin{equation}
    H_{\text{eff}}^{\text{tri}} = \lim_{V \rightarrow \infty} \mathcal{P}_{\text{tri}} H^{(2)}_{\text{eff}} \mathcal{P}^{\dagger}_{\text{tri}} = \lim_{V \rightarrow \infty}
    \begin{pmatrix}
        1 & 0 & 0 & 0 \\
        0 & \frac{1}{\sqrt{2}} & \frac{1}{\sqrt{2}} & 0 \\
        0 & 0 & 0 & 1
    \end{pmatrix}
    H^{(2)}_{\text{eff}}
     \begin{pmatrix}
        1 & 0 & 0 \\
        0 & \frac{1}{\sqrt{2}} & 0\\
        0 & \frac{1}{\sqrt{2}} & 0 \\
        0 & 0 & 1
    \end{pmatrix} = 
    \lim_{V \rightarrow \infty} (4V - i \gamma^g_{\uparrow}/4) + h(k),
\end{equation}
where $h(k)$ is a $3\times 3$ matrix within triplet subspace.
Its eigenvalues are given by $\epsilon_0(k)=0;\epsilon_{\pm} = \pm \epsilon_k =\pm \sqrt{(i \gamma^g_{\uparrow}/4 + 2 \sin{k})^2 + 4 \, (3/2 + \cos{k})^2}$. 
Finally, in the strong interaction limit, the return function simplifies to $g(k,t) = |\langle\Psi_0|e^{-i H_{\text{eff}}^{(2)}(k) t}|\Psi_0\rangle|^2 = |\langle\Psi_0|\mathcal{P}^{\dagger}_{\text{tri}} \mathcal{P}_{\text{tri}} e^{-i H_{\text{eff}}^{(2)}(k) t} \mathcal{P}^{\dagger}_{\text{tri}} \mathcal{P}_{\text{tri}}|\Psi_0\rangle|^2 = |\langle\Psi_0|\mathcal{P}^{\dagger}_{\text{tri}} e^{-i h(k) t} \mathcal{P}_{\text{tri}}|\Psi_0\rangle|^2= |\langle \Psi_0|\mathcal{P}^{\dagger}_{\text{tri}} e^{-i h(k) t} \mathcal{P}_{\text{tri}}|\Psi_0\rangle|^2$. 

To identify the critical times at which DQPTs occur, we analytically determine the Fisher zeros. 
Fisher zeros are firstly introduced in equilibrium quantum phase transitions, which refer to the complex zeros of the partition function with respect to an external control parameter, such as temperature or a magnetic field~\cite{YangLee1952}.
In the context of dynamical quantum phase transition, Fisher zeros correspond to the complex zeros of return function $g(t) = \langle \Psi_0|e^{- i H t}|\Psi_0\rangle$. 
A DQPT occurs when these Fisher zeros cross the real-time axis in the complex-$t$ plane. 
Now we calculate the quantity 
\begin{equation}
\begin{split}
    \langle\Psi_0|\mathcal{P}^{\dagger}_{\text{tri}} e^{-i h(k) t} \mathcal{P}_{\text{tri}}|\Psi_0\rangle & = \langle \psi_0|e^{-i h(k) t}|\psi_0\rangle \\ 
    & = \langle \psi_0|\varphi^R_0\rangle \langle\varphi^L_0|\psi_0\rangle + e^{-i\epsilon_+ t} \langle \psi_0|\varphi^R_+\rangle \langle\varphi^L_+|\psi_0\rangle + e^{-i\epsilon_- t} \langle \psi_0|\varphi^R_-\rangle \langle\varphi^L_-|\psi_0\rangle \\
    & = c_0(k) + e^{-i \epsilon_k t} c_+(k) + e^{i \epsilon_k t} c_-(k),
\end{split}
\end{equation}
where the superscript $R$ and $L$ represents the right and left eigenvectors, respectively. 
The fisher zeros can be analytically solved as 
\begin{equation}
    t_{c,n}(k) = \frac{1}{\epsilon_k} \left(2\pi n  - i \log\left[\frac{-c_0+\sqrt{2c_0(k) + (c_+(k) - c_-(k))^2 - 1}}{1+ c_+(k) -c_0(k) - c_-(k)}\right]\right),
\end{equation}
where $n$ denotes different branch of solution $t$. 
As $k$ spans from $0$ to $2\pi$, $t_{c,n}(k)$ generates a set of algebraic curves in the complex-$t$ plane, as shown in Fig.~\ref{fig:S1}(b). 
The intersections between $t_{c,n}(k)$ and real axis correspond to the critical times at which DQPTs occur, which align with the cusps illustrated in Fig.~\ref{fig:S1}(a). 

So far, we demonstrate that in the extreme cases with $V=0$ and $V=\infty$, DQPTs occur and can be fully captured by the corresponding post-quench effective Hamiltonians. 
For finite interaction strength $V$, DQPTs persist and continuously interpolate between these two limiting cases. 

Finally, we perturbatively introduce particle loss on spin-$\uparrow$ $L_x^l = \sqrt{\gamma_{\uparrow}^l}c_{x,A,\uparrow}$ into the post-quench Liouvillian superoperator. 
According to the theorem, the presence of both gain and loss leads to the suppression of DQPTs, which is confirmed numerically in Fig.~\ref{fig:S1}(c).
For a fixed momentum $k$, the full Liouvillian takes the block tridiagonal form
\begin{equation}
    L(k) = \begin{pmatrix}
    L_0^{(0)} & L_d^{(0\leftarrow 1)} & 0 & 0 & 0 \\ 
    L_u^{(1\leftarrow 0)} & L_0^{(1)} & L_d^{(1\leftarrow 2)} & 0 & 0 \\
    0 & L_u^{(2\leftarrow 1)} & L_0^{(2)} & L_d^{(2\leftarrow 3)} & 0 \\
    0 & 0 & L_u^{(3\leftarrow 2)} & L_0^{(3)} & L_d^{(3\leftarrow 4)} \\
    0 & 0 & 0 & L_u^{(4\leftarrow 3)} & L_0^{(4)}
    \end{pmatrix},
\end{equation}
where the block-diagonal entries are given by $L_0^{(n)} = -i (H^{(n)}_{\text{eff}}\otimes I - I \otimes (H^{(n)}_{\text{eff}})^{\ast})$, with $n$ denoting the occupation number sector at fixed $k$. 
We omit the detailed derivations here, as they follow the same steps as the two-band example in Section~\ref{SecII}.
In summary, we have applied the theorem to a solvable interacting system and confirmed its validity. 

\section{IV.~~Monte-Carlo wavefunction method}\label{SecIV}
When studying Markovian open quantum systems whose third quantized Lindbladian is not quadratic, analytical solution becomes unattainable and numerical methods need to be applied.
If we directly solve the Lindblad master equation, we are essentially diagonalizing a Liouvillian matrix of dimension $(\dim \mathcal{H})^2$, which is computationally expensive.
In this section, we introduce the quantum trajectory method~\cite{Molmer1992PRL}, which converts the Lindblad evolution of density matrix $\rho$ to a stochastics differential equation of wavefunction $\ket{\psi(t)}$ satisfying $\rho(t)=\mathbb{E}[\ket{\psi(t)}\bra{\psi(t)}]$.
Using Monte-Carlo method, we simulate an ensemble of wavefunctions in parallel, thus efficiently evaluating the DQPT rate functions by averaging over the ensemble.

Consider the following quantum channel under continuous measurements: within a infinitesimal time interval $(t,t+\delta t)$, the channel is subject to a set of measurement operators $\{M_0,M_1,\cdots,M_n\}$ given by
\begin{equation}
    M_0=I-i H_{\text{eff}} \delta t,\quad M_\mu = L_\mu \sqrt{\delta t}, \mu=1,2,\cdots,n,
\end{equation}
where $H_{\text{eff}}=H-\frac{i}{2}\sum_\mu L^\dagger_\mu L_\mu$ so that the $M_i,i=0,1,\cdots,m$ satisfy the normalization condition of positive operator-valued measure (POVM):
\begin{equation}\label{eq:povm_norm}
    \sum_{j=0}^{n}M_j^\dagger M_j=M_0^\dagger M_0+\sum_{\mu} M_{\mu}^\dagger M_\mu=I+i(H^\dagger_{\text{eff}}-H_{\text{eff}})\delta t+\sum_{\mu} L_{\mu}^\dagger L_{\mu}\delta t=I.
\end{equation}
Here, $M_0$ corresponds to the non-unitary time evolution under the effective non-Hermitian Hamiltonian $H_{\text{eff}}$, while $M_{\mu}$ corresponds to quantum jumps. The non-unitary evolution can be seen as the deterministic ``drift" process, while the quantum jumps are the stochastic ``diffusion" processes.
Suppose the wavefunction at time $t$ is given by $\ket{\psi(t)}$, then during $(t,t+\delta t)$, the possibility of deterministic non-unitary evolution and random jumps are respectively given by:
\begin{equation}
   p_0=\mel{\psi(t)}{M_0^\dagger M_0}{\psi(t)},\quad p_\mu=\mel{\psi(t)}{M_\mu^\dagger M_\mu}{\psi(t)}=\mel{\psi(t)}{L_\mu^\dagger L_\mu}{\psi(t)}\delta t,\ \mu=1,\cdots,n.
\end{equation}
From Eq.~\eqref{eq:povm_norm}, we have $\sum_{j=0}^n p_j=1$. With probability $p_j$, the wavefunction evolves to the state $\ket{\psi_j(t+\delta t)}$ given by
\begin{equation}
    \ket{\psi_j(t+\delta t)}=\frac{M_j\ket{\psi(t)}}{\sqrt{\mel{\psi(t)}{M_j^\dagger M_j}{\psi(t)}}},\ j=0,1,\cdots,n.
\end{equation}
Therefore, the density matrix at $t+\delta t$ with the statistical ensemble is given by
\begin{equation}
    \begin{aligned}
        \rho(t+\delta t)&=\mathbb{E}[\ket{\psi(t+\delta t)}\bra{\psi(t+\delta t)}]\\
        &=p_0\mathbb{E}[\ket{\psi_0(t+\delta t)}\bra{\psi_0(t+\delta t)}]+\sum_{\mu=1}^n p_\mu \mathbb{E}[\ket{\psi_\mu(t+\delta t)}\bra{\psi_\mu(t+\delta t)}]\\
        &=(I-i H_{\text{eff}}\delta t)\mathbb{E}[\ket{\psi(t)}\bra{\psi(t)}](I+i H_{\text{eff}}^\dagger \delta t)+\sum_{\mu=1}^n L_\mu \mathbb{E}[\ket{\psi(t)}\bra{\psi(t)}] L_\mu^\dagger \delta t\\
        &=\rho(t)-i\left[H_{\text{eff}}\rho(t)-\rho(t)H^\dagger_{\text{eff}}\right]\delta t+\sum_{\mu=1}^n L_\mu\rho(t)L_\mu^\dagger\delta t +o(\delta t^2).
    \end{aligned}
\end{equation}
Then, we recover the Lindblad master equation:
\begin{equation}
    \frac{\dd \rho(t)}{\dd t}=\lim_{\delta t\to 0}\frac{\rho(t+\delta t)-\rho(t)}{\delta t}=-i\left[H_{\text{eff}}\rho(t)-\rho(t)H^\dagger_{\text{eff}}\right]+\sum_{\mu=1}^n L_\mu\rho(t)L_\mu^\dagger.
\end{equation}

In the last section of the main text, we studied a 1D SSH model with two-body coherent jumps where $L\sim c_A c_B$ and $c_A^\dagger c_B^\dagger$ so that the $H_{\text{eff}}$ contains an imaginary interaction term.
The Lindbladian dynamics of this system is calculated numerically via the Monte-Carlo wavefunction method, where we let $M$ copies of wavefunctions evolve stochastically in parallel. 
For each trajectory $m=1,\ldots,M$, the wavefunction $\ket{\psi^{(m)}(t)}$ is propagated according to the stochastic evolution described above. The density matrix at time $t$ is then approximated by the ensemble average:
\begin{equation}
    \rho(t) = \mathbb{E}[\ket{\psi(t)}\bra{\psi(t)}] \approx \frac{1}{M} \sum_{m=1}^M |\psi^{(m)}(t)\rangle \langle\psi^{(m)}(t)|.
\end{equation}
Physical observables $\hat{\mathcal{O}}$ are evaluated by averaging over these trajectories. 
\begin{equation}
    \hat{\mathcal{O}}(t) = \Tr{\rho(t)\hat{\mathcal{O}}}=\frac{1}{M}\sum_{m=1}^M \langle \psi^{(m)}(t)|\hat{\mathcal{O}}|\psi^{(m)}(t)\rangle.
\end{equation}
In the following we list all the key steps in numerical realization:
\begin{enumerate}
    \item Initialize the system, set $\ket{\psi_0}$ as the half-filled ground state of the pre-quench Hamiltonian $H_0$. Prepare the post-quench effective non-Hermitian Hamiltonian $H_{\text{eff}}$ as well as the jump operators ${L_{\mu}}$.
    \item Create $M$ copies of $\ket{\psi_0}$ and evolve each one in time using discrete time steps $t_{j+1}-t_j=\Delta t$. Suppose at time $t=t_j$, the $m$-th copy of wavefunction is $|\psi^{(m)}(t_j)\rangle$, then calculate the probabilities:
    \[p_\mu=\langle \psi^{(m)}(t_j)|L_\mu^\dagger L_\mu|\psi^{(m)}(t_j)\rangle\Delta t,\ \mu=1,\cdots,n,\quad p_0=1-\sum_{\mu=1}^{n}p_\mu.\]
    Define a cumulative quantity $q_j=\sum^{j}_{s=0}p_s$ for $j=0,1,\cdots,n$. 
    \item Evolve the wavefunction with the calculated probabilities $p_j$. Generate a random number $u$ with an uniform probability density within $u\in(0,1)$. Compare $u$ with the elements in the set $Q=\{q_0,q_1,\cdots,q_n\}$:
    \begin{itemize}
        \item If $u<q_0$, then, the wavefunction undergoes non-unitary evolution, define the unnormalized
        \[|\widetilde{\psi}^{(m)}(t_{j+1})\rangle=(I-i H_{\text{eff}}\Delta t)|\psi^{(m)}(t_j)\rangle\]
        \item If $u>q_0$ and $q_\nu=\min\{q\in Q:\ u<q\}$, then a quantum jump occurs and we have
        \[|\widetilde{\psi}^{(m)}(t_{j+1})\rangle=L_\nu|\psi^{(m)}(t_j)\rangle\]
    \end{itemize}
    \item Normalize the wavefunction:
    \[|\psi^{(m)}(t_{j+1})\rangle=\frac{|\widetilde{\psi}^{(m)}(t_{j+1})\rangle}{\sqrt{\langle\widetilde{\psi}^{(m)}(t_{j+1})|\widetilde{\psi}^{(m)}(t_{j+1})\rangle}}\]
    Use this wavefunction to calculate the probabilities for the next time step $t_{j+1}\to t_{j+2}$, repeating steps 2-3.
    \item Calculate the Loschmidt rate function by doing ensemble average:
    \[ g(t) =\frac{1}{M}\sum_{m=1}^M\abs{\langle\psi(0)|\psi^{(m)}(t)\rangle}^2.\]
\end{enumerate}
This approach significantly reduces computational complexity compared to direct integration of the full Lindblad equation, especially for large Hilbert spaces.


%


\end{document}